\documentclass[twocolumn,showpacs,preprintnumbers,amsmath,amssymb]{revtex4}

\usepackage{graphicx}
\usepackage{bm}

\begin{document}

\title{Generation and control of resonance states in crossed magnetic
and electric fields}

\author{K. Krajewska and J. Z. Kami\'nski}

\affiliation{Institute of Theoretical Physics, Warsaw University,
Ho\.za 69, 00-681 Warszawa, Poland}


\begin{abstract}
A two-dimensional electron system interacting with an impurity and
placed in crossed magnetic and electric fields is under
investigation. Since it is assumed that an impurity center
interacts as an attractive $\delta$-like potential a
renormalization procedure for the retarded Green's function has to
be carried out. For the vanishing electric field we obtain a close
analytical expression for the Green's function and we find one
bound state localized between Landau levels. It is also shown by
numerical investigations that switching on the electric field new
long-living resonance states localized in the vicinity of Landau
levels can be generated.
\end{abstract}

\maketitle


\section{Introduction}
\label{sec:Introduction}

The quantum mechanical motion of electrons in two-dimensional
space, subject to crossed magnetic and electric fields and in the
presence of a single impurity described by a zero-range potential,
provides a basis for the explanation of the integer quantum Hall
effect \cite{P81,PJ82,JP84,PC91,CC98,GM99,HL99}. Until now in all
considerations the electric field has been treated as a small
perturbation. The purpose of the present work is to reconsider
these investigations and to discuss some novel phenomena which
appear for nonperturbative electric fields.

In our approach an attractive impurity is modelled as a zero-range
potential center, as has been already exploited in
\cite{PC91,CC98}. It turns out that in order to define such a
point interaction one needs to renormalize the strength of the
$\delta$-function and that such a renormalized potential is
capable of supporting one localized state per Landau level. This
finding also corresponds to the result obtained by Gyger and
Martin \cite{GM99}. Investigating the case of a weak electric
field they have proven that all localized states change into
resonances with lifetimes varying with the electric field in a
Gaussian way. With reference to these papers we shall focus our
attention on the opposite regime, i.e., when the electric fields
is strong. Our aim is to answer the following questions:
\begin{enumerate}
\item
how many resonance states exist per one Landau level for a non-vanishing
electric field?
\item
what is a maximum electric field above which resonance states of a
given lifetime cannot exist?
\end{enumerate}
We shall see that apart from the well-known localized states lying
between Landau levels \cite{PC91,CC98,GM99} \textit{new resonance
states} are created by a strong electric field, the lifetime of
which can be comparable to or larger than typical times
encountered in solid state physics.

Throughout this paper we use units in which $\hbar=1$.

\section{Attractive point interaction}
\label{sec:math}

The two-dimensional system we are concerned with is governed by
the hamiltonian which, for convenience, we decompose into two
parts, $H_0$ related to the physical situation for which the
Green's function is known and a $\delta$-like potential with the
strength given by the so-called bare coupling constant $\lambda$,
\begin{equation}
H=H_0+\lambda \delta^{(2)} (\bm{r}) \, . \label{e2.1}
\end{equation}
Such a model has been extensively studied in the literature
\cite{AGHH88,B75,FFR90,FFR91,MFSF00,W91} proving its usefulness in
many branches of physics. It is, however, important to realize
that the model is not free from difficulties. To be more precise,
it exhibits divergencies which have to be tempered by
renormalizing the bare coupling constant $\lambda$. Therefore, let
us address this problem and show how to remove all divergent terms
by considering the retarded Green's function for the hamiltonian
$H$.

Our starting point is the Lippmann-Schwinger equation which, in
this particular case, adopts the form
\begin{eqnarray}
G(\bm{r},\bm{r}';E) = 
G_0^{(\mathrm{reg})}(\bm{r},\bm{r}';E) \nonumber \\ 
+ \int \mathrm{d}^2\bm{r}''
G_0^{(\mathrm{reg})}(\bm{r},\bm{r}'';E)\lambda
\delta^{(2)} (\bm{r}'') G(\bm{r}'',\bm{r}';E) \, ,
\label{e2.2}
\end{eqnarray}
where the regularized Green's function for the hamiltonian $H_0$,
denoted by $G_0^{(\mathrm{reg})}(\bm{r},\bm{r}';E)$, is
assumed to be known. Now performing the foregoing integral over
$\text{d}^2\bm{r}''$ and making some algebraic manipulations
we arrive at the retarded Green's function for our model
\begin{eqnarray}
G(\bm{r},\bm{r}';E)=
G_0^{(\text{reg})}(\bm{r},\bm{r}';E) \nonumber\\
+ \frac{G_0^{(\text{reg})}(\bm{r},\bm{0};E)
G_0^{(\text{reg})}(\bm{0},\bm{r}';E)}
{\lambda^{-1}-G_0^{(\text{reg})}(\bm{0},\bm{0};E)} \, .
\label{e2.3}
\end{eqnarray}
It appears that $G_0^{(\text{reg})}(\bm{0},\bm{0};E)$ is
the only term which needs to be regularized in order to avoid
divergencies. Let us explain how to cope with them analyzing the
simplest case of a free electron with an effective mass $m^*$
interacting only with the $\delta$-like potential. In our approach
we choose a particular regularization scheme which we call the
proper time regularization. We expect that although the bare
coupling constant $\lambda$ depends on the regularization scheme
chosen, the observable quantities, like poles of the full Green's
function in the complex energy plane, cannot depend on this
scheme. Indeed, we shall demonstrate that our method leads to the
same equation for energies of localized states in the magnetic
field as in \cite{CC98}.

For a free particle the retarded Green's function in the so-called
Fock-Schwinger proper time represantation \cite{IZ80} has the form
\begin{eqnarray}
G_0^{(\text{reg})}(\bm{r},\bm{r}';E)
=-\text{i}\int_{-\text{i}\sigma}^{\infty} \text{d}t
\text{e}^{\text{i}(E+\text{i}\varepsilon)t} \nonumber\\
\int \frac{\text{d}^2\bm{k}}{(2\pi)^2} \exp\biggl
(\text{i}\bm{k}\cdot (\bm{r} -\bm{r}')
-\text{i}\frac{\bm{k}^2}{2m^*}t \biggr ) \, , \label{e2.4}
\end{eqnarray}
where the singularity for $\bm{r}=\bm{r}'=\bm{0}$ is
avoided by modifying the integration contour over the proper time
$t$ in such a way that it lies just below the real axis (this
means that $\sigma$ is a small positive parameter). Naturally, the
retarded Green's function for non-vanishing $\bm{r}$ or
$\bm{r}'$ is recovered by putting $\sigma=0$. Such a
prescription we call the proper time regularization. Now, it is
straightforward to show that the denominator in eq. (\ref{e2.3})
reads
\begin{equation}
D(E)=
\lambda^{-1}+\frac{m^*}{2\pi}\int_{-\text{i}\sigma}^{\infty}
\text{d}t \text{e}^{\text{i}(E+\text{i}\varepsilon)t}
t^{-1} \, . \label{e2.5}
\end{equation}
It is also clear that the integral above diverges logarithmically
for small values of $t$ when $\sigma=0$. This divergence, however,
can be absorbed into the bare coupling constant $\lambda$ giving
the finite expression for the full Green's function even for
$\sigma$ equal to 0. Indeed, performing the proper time
integration by parts we arrive at
\begin{equation}
D(E)=\lambda_R^{-1}+\frac{m^*}{2\pi} \biggl ( \ln(\text{i})
-\text{i}E\int_0^{\infty} \text{d}t
\text{e}^{\text{i}(E+\text{i}\varepsilon)t}\ln(t/t_0) \biggr
) \, , \label{e2.6}
\end{equation}
where $t_0$ is an arbitrary positive parameter of the same
dimensionality as $t$, whereas $\lambda_R$ is the renormalized
coupling constant defined by the relation
\begin{equation}
\lambda_R^{-1}=\lim_{\sigma\rightarrow 0} \Bigl (
\lambda^{-1}-\frac{m^*}{2\pi}\ln \frac{\sigma}{t_0}  \Bigr ) \, .
\label{e2.7}
\end{equation}
Let us remark that in order to temper a singular part in the
formula above the bare coupling constant $\lambda$ has to be
negative. This means, in fact, that in two dimensions there cannot
exist a repulsive point interaction. Having this in mind we shall
restrict ourselves only to the case with an attractive point
interaction for which, after performing the remaining integral, we
have
\begin{equation}
D(E)=\lambda_R^{-1}+\frac{m^*}{2\pi}\ln\biggl (
\frac{\text{e}^{-\gamma}}{-Et_0} \biggr ) \, , \label{e2.8}
\end{equation}
where $\gamma$ is the Euler's constant. This equation shows that
independently of numerical values of $\lambda_R$ and $t_0$ there
exists $E$ such that the denominator of the full Green's function
vanishes, i.e., the two-dimensional zero-range potential always
supports a bound state. Denoting the energy of this state by $E_B$
we end up with the following expression for the denominator
$D(E)$,
\begin{equation}
D(E)=\frac{m^*}{2\pi}\ln\biggl (\frac{E_B}{E} \biggr ) \, .
\label{e2.9}
\end{equation}
Let us add in closing that there still remains a free parameter of
the theory, i.e., the energy of a bound state $E_B$. Nevertheless,
it has now the clear physical meaning in the sense that its value
can be measured, contrary to the artificial and physically unclear
parameters $t_0$ and $\lambda_R$.

\section{Green's function in external fields}
\label{sec:Green'sfunction}

In this section we consider the two-dimensional system composed of
an electron and $\delta$-like impurity placed in the external
magnetic and electric fields for which the hamiltonian $H_0$ can
be written as
\begin{equation}
H_0=\frac{1}{2m^*}\Biggl[ \Bigl(
-\text{i}\partial_x+\frac{e\cal{B}}{2}y\Bigr )^2+
\Bigl(-\text{i}\partial_y-\frac{e\cal{B}}{2}x\Bigr )^2\Biggr
]-e{\cal E}x \, , \label{e3.1}
\end{equation}
where the electric $\bm{\mathcal{E}}$ and magnetic
$\bm{\mathcal{B}}$ fields are directed towards $x$ and $z$
axes, respectively, whereas the vector potential is taken in the
circular gauge,
$\bm{A}=\frac{1}{2}\bm{\mathcal{B}}\times\bm{r}$. As
we have already pointed out, the problem of finding energies of
impurity-induced resonance states consists in searching for zeros
of the denominator in eq. (\ref{e2.3}). Therefore, what we have to
know is the regularized Green's function for the hamiltonian
$H_0$. Applying the proper time regularization scheme discussed
above we arrive at the regularized retarded Green's function
$G_0^{(\text{reg})}$,
\begin{eqnarray}
G_0^{(\text{reg})}(\bm{r},\bm{r}';E) = 
-\frac{m^*\omega}{4\pi}\int_{-\text{i}\sigma}^{\infty}
\text{d}t \frac{\text{e}^{\text{i}(E+\text{i}\varepsilon)t}}
{\sin\frac{\omega t}{2}} \nonumber\\
\times \exp\Biggl[ \frac{\text{i}m^*\omega}{4}\bigl( (x-x')^2 
 + (y-y')^2\bigr )\cot\frac{\omega t}{2} \nonumber\\
+\frac{\text{i}m^*\omega}{2}(xy'-x'y)
+\frac{\text{i}e{\cal E}t}{2}(x+x') \nonumber\\
 + \frac{\text{i}e\cal{E}}{\omega}\Bigl( \frac{\omega
t}{2} \cot\frac{\omega t}{2}-1\Bigr ) \Bigl(
-y+y'+\frac{e{\cal E}t}{2m^*\omega}\Bigr )\Biggr ] \, ,
\label{e3.2}
\end{eqnarray}
and the denominator $D(E)$,
\begin{eqnarray}
D(E)=\lambda^{-1}+\frac{m^*\omega}{4\pi}\int_{
-\text{i}\sigma}^{\infty} \text{d}t
\frac{\text{e}^{\text{i}(E+\text{i}\varepsilon)t}}
{\sin\frac{\omega t}{2}} \nonumber\\
\times \exp\Biggl[
\frac{\text{i}t}{2m^*}\Bigl( \frac{e\cal{E}}{\omega}\Bigr
)^2 \Bigl( \frac{\omega t}{2}\cot\frac{\omega t}{2}-1\Bigr )\Biggl
] \, , \label{e3.3}
\end{eqnarray}
where $\omega=|e|{\cal B}/m^*$ is the cyclotron frequency.
Comparing this formula with eq. (\ref{e2.5}) one can notice that
in both these expressions the integrals over $t$ diverge for small
arguments exactly in the same way. Hence, we gather that now it is
also possible to temper the divergent term by renormalizing the
coupling constant $\lambda$. At this point let us introduce new
dimensionless units. As a typical length scale for the problem we
take $1/\sqrt{m^{*}\omega}$, whereas the dimensionless scaled
electric field and energy are
$\tilde{\cal E}=|e|{\cal E}/\sqrt{m^* \omega^3}$ and
$\tilde{E}=2 E/\omega$, respectively. Additionally, choosing a new
variable $s=\omega t/2$ we find that
\begin{eqnarray}
D(\tilde{E})=\frac{m^*}{2\pi}\Biggl [\ln\biggl(
\frac{\tilde{E}_B}{\tilde{E}}\biggr )+ \int_0^{\infty}\text{d}s
\text{e}^{\text{i}(\tilde{E}+ \text{i}\varepsilon)s} \nonumber\\
\biggl(\frac{\exp\bigl( \text{i} {\tilde{\cal E}}^2 s(s\cot
s-1)\bigr )}{\sin s}-\frac{1}{s}\biggr )\Biggr ] \, . \label{e3.4}
\end{eqnarray}
At first glance it might seem that the integral above is divergent
for $s=n\pi$, where $n=1, 2,..$. It is not, however, the case.
Indeed, analyzing the integrand in the close vicinity of these
points one can prove the convergence of the foregoing integral.
Having this in mind we shall discuss below the special case when
the electric field is switched off.
\begin{figure}
\begin{center}
\includegraphics[width=7.5cm]{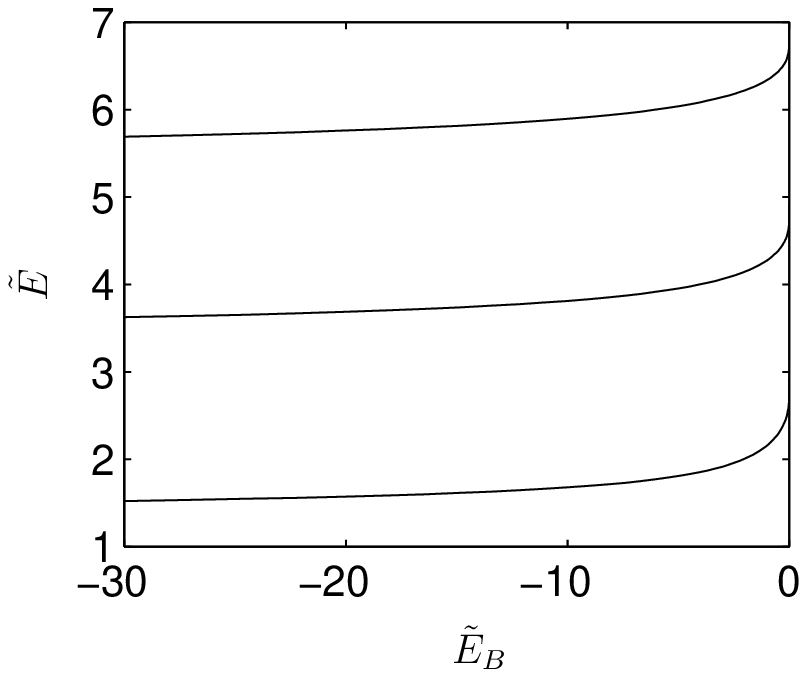}
\end{center}
\caption{Impurity-induced localized states for the vanishing
electric field} \label{fig1}
\end{figure}
\begin{figure}
\begin{center}
\includegraphics[width=7.5cm]{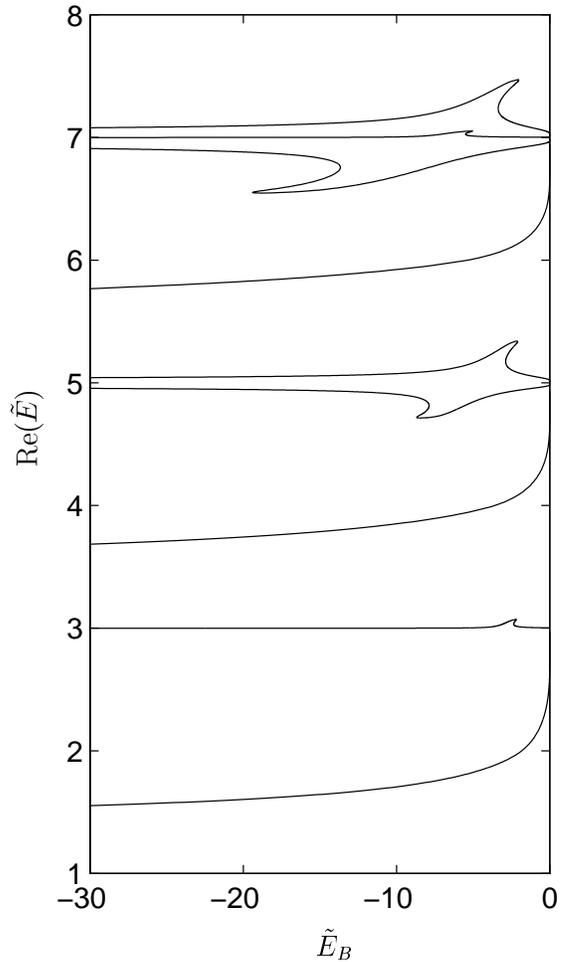}
\end{center}
\caption{Real part of resonance states as the function of
$\tilde{E}_B$ for $\text{Im}(\tilde{E})=10^{-4}$.} \label{fig2}
\end{figure}
\begin{figure}
\begin{center}
\includegraphics[width=7.5cm]{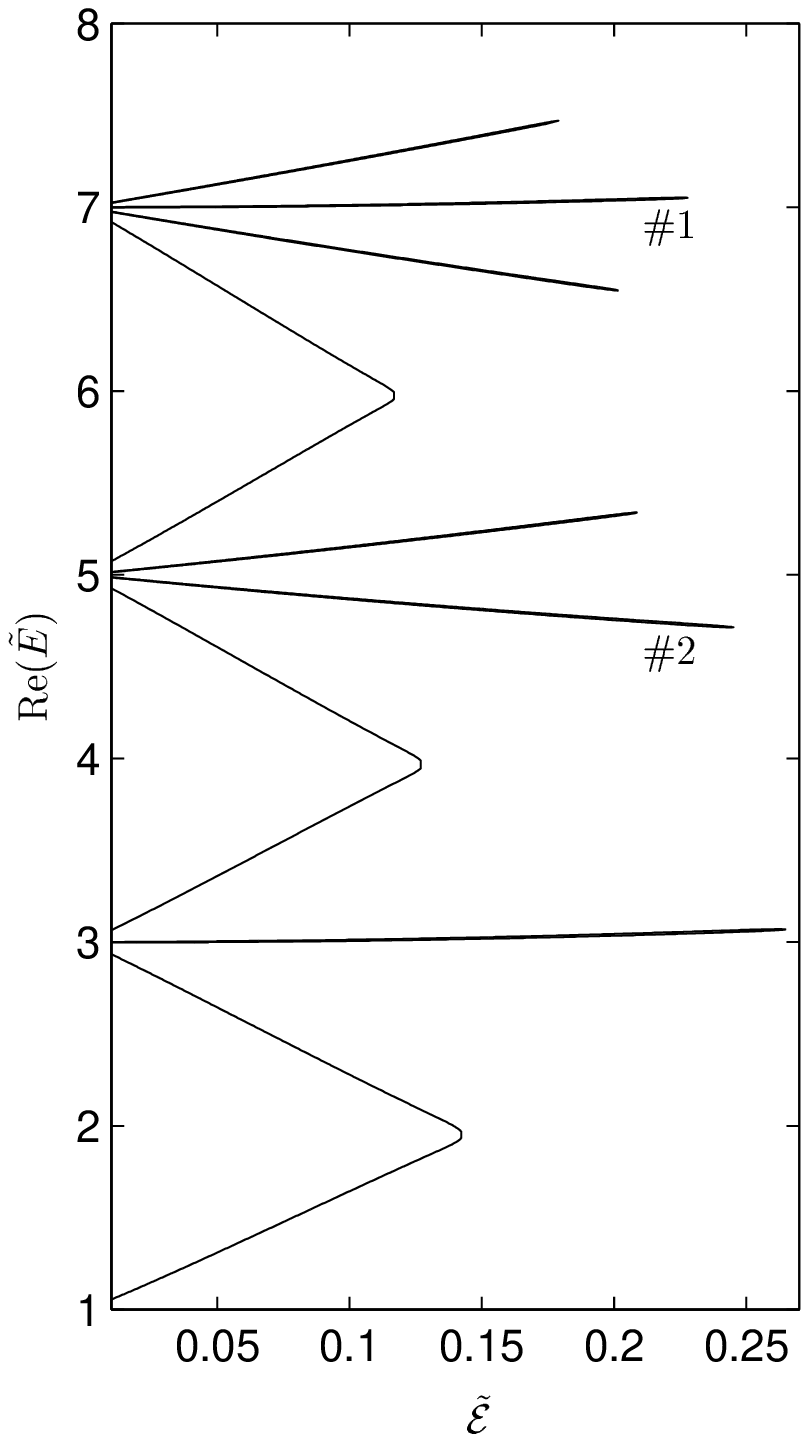}
\end{center}
\caption{Real part of resonance states as the function of
$\tilde{\cal{E}}$ for $\text{Im}(\tilde{E})=10^{-4}$.}
\label{fig3}
\end{figure}
\begin{figure}
\begin{center}
\includegraphics[width=7.5cm]{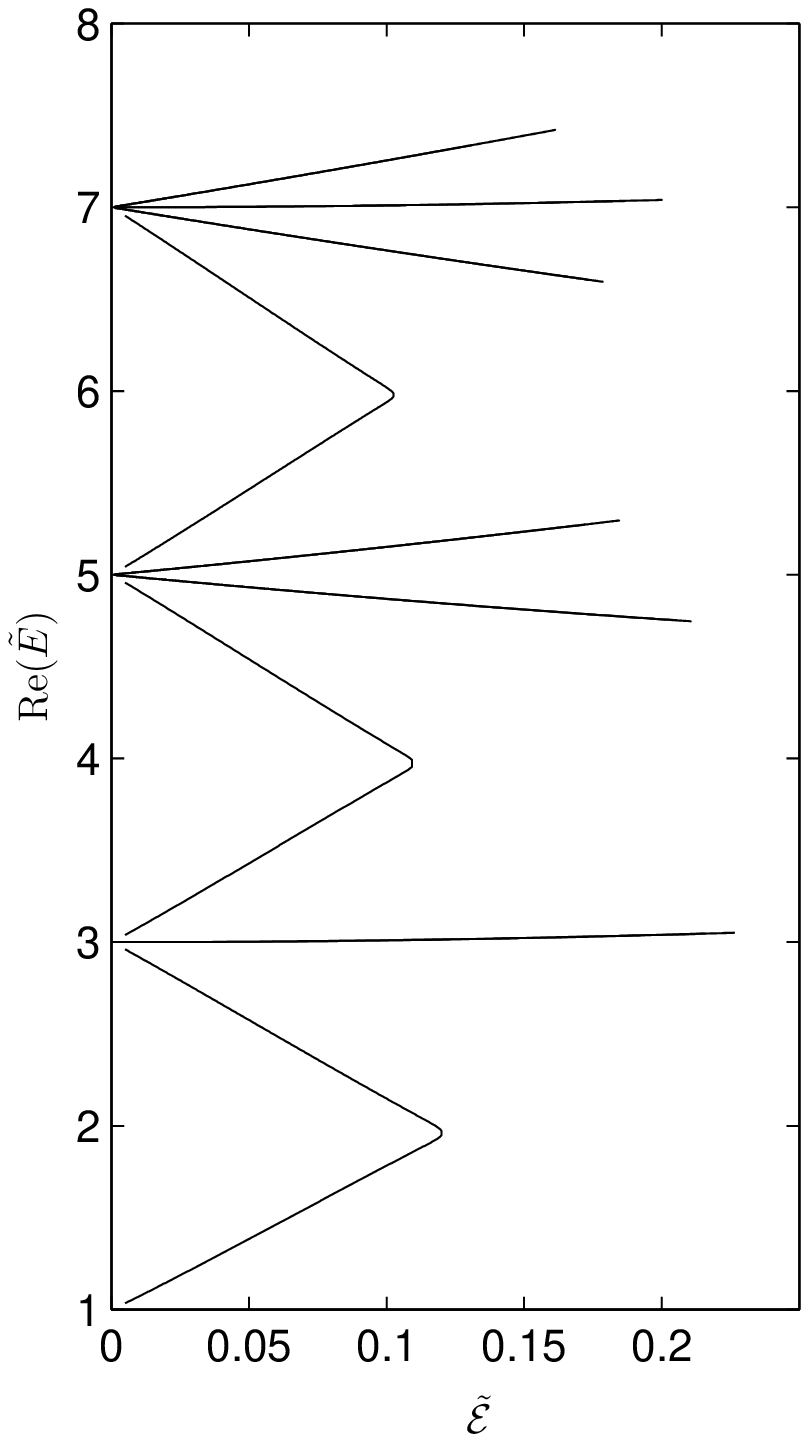}
\end{center}
\caption{Real part of resonance states as the function of
$\tilde{\cal{E}}$ for $\text{Im}(\tilde{E})=10^{-6}$.}
\label{fig4}
\end{figure}

\section{Localized states in the external magnetic field}
\label{sec:localizedstates}

The problem without an electric field has been already examined by
Cavalcanti and de Carvalho in \cite{CC98}. However, we have found
it interesting to derive their equation for energies of the
impurity-induced localized states using the proper time
regularization scheme. Our point of departure is the function
defined by eq. (\ref{e3.4}) with $\tilde{\cal E}=0$,
\begin{equation}
D(\tilde{E})=\frac{m^{*}}{2\pi}\Biggl[ \ln\biggl(
\frac{\tilde{E}_B}{\tilde{E}}\biggr ) +\int_0^{\infty}\text{d}s
\text{e}^{\text{i}(\tilde{E}+\text{i}\varepsilon)s}
\biggl(\frac{1}{\sin s}-\frac{1}{s}\biggr) \Biggr] \, .
\label{e3.5}
\end{equation}
In order to treat these two integrals separately we regularize
them first. This allows us to integrate them by parts and put
finally the regularization parameter $\sigma$ equal to 0, keeping
in mind, however, that the integration contour over $s$ lies just
below the real axis,
\begin{eqnarray}
D(\tilde{E})=\frac{m^{*}}{2\pi}\Biggl[ \gamma+\ln(2\tilde{E}_B)
-\frac{\text{i}\pi}{2} \nonumber\\
-\text{i}\tilde{E} \int_{0}^{\infty}\text{d}s
\text{e}^{\text{i}(\tilde{E}+\text{i}\varepsilon)s} \ln\tan
(s/2) \Biggr] \, . \label{e3.6}
\end{eqnarray}
Next, let us expand $\ln\tan(s/2)$ into series of
$\text{e}^{-\text{i}s}$. This is allowed because the
inequality $|\text{e}^{-\text{i}s}|<1$ holds for $s$
possessing a negative imaginary part, which is guaranteed by our
choice of a regularization scheme. Hence, inserting into eq.
(\ref{e3.6}) the series expansion
\begin{equation}
\ln\tan(s/2)=-\frac{\text{i}\pi}{2}-2\sum_{n=0}^{\infty}
\frac{\text{e}^{-\text{i}(2n+1)s}}{2n+1} \, , \label{e3.7}
\end{equation}
and carrying out all integrations term by term, we arrive at
\begin{eqnarray}
D(\tilde{E})=\frac{m^{*}}{2\pi}\Biggl[
\gamma+\ln(2\tilde{E}_B)-\text{i}\pi \nonumber\\
+ 2\tilde{E}\sum_{n=0}^{\infty}\frac{1}{(2n+1)(2n+1-\tilde{E})}
\Biggr] \, . \label{e3.8}
\end{eqnarray}
Finally, relating the sum above to a digamma function
$\psi(z)$, we find that the eigenenergies of our system are
zeros of the following function
\begin{equation}
D(\tilde{E})=\frac{m^{*}}{2\pi}\Biggl[ \ln\biggl(
\frac{|\tilde{E}_B|}{2}\biggr)
-\psi\biggl(\frac{1-\tilde{E}}{2}\biggr) \Biggr] \, , \label{e3.9}
\end{equation}
which is adequate to the result derived in \cite{CC98} with the
help of another regularization scheme. Zeros of $D(\tilde{E})$
determine the dependence of $\tilde{E}$ on $\tilde{E}_B$ for those
localized states which are extracted from the Landau levels by the
zero-range interaction. It is shown in Fig. \ref{fig1} that only
one state is extracted per Landau level, which is understandable,
since in the degenerate Landau level there is only one state the
wavefunction of which does not vanish at the origin, i.e., the
cylindrically symmetric state \cite{BBCK}. The situation changes
if the electric field is present. It mixes this state with those
Landau states which explicitly depend on the polar angle. For
small electric fields such a mixing is also small and the effect
of a point interaction can be neglected. However, for strong
electric fields this effect becomes considerable as we are going
to show this below.
\begin{figure}
\begin{center}
\includegraphics[width=7.5cm]{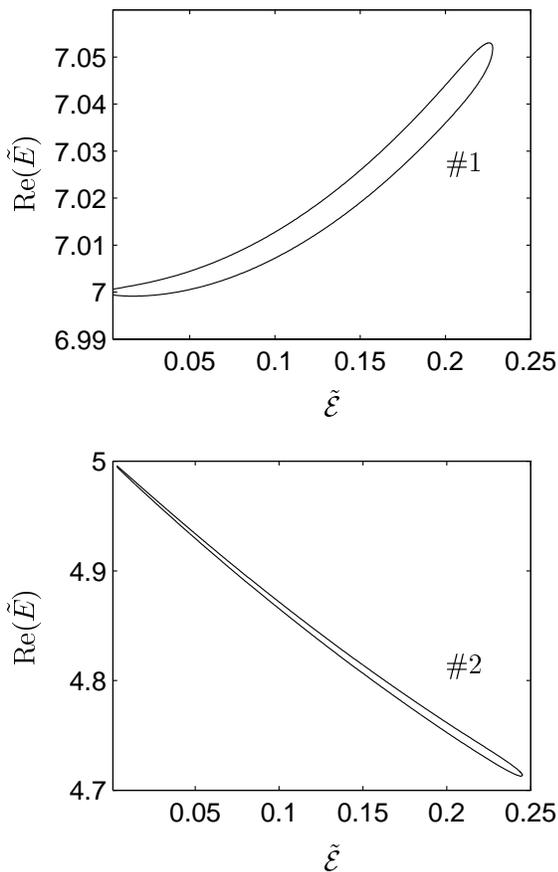}
\end{center}
\caption{Two resonance states labeled by $\#1$ and $\#2$ in Fig. 
\ref{fig3}.} \label{fig5}
\end{figure}
\begin{figure}
\begin{center}
\includegraphics[width=7.5cm]{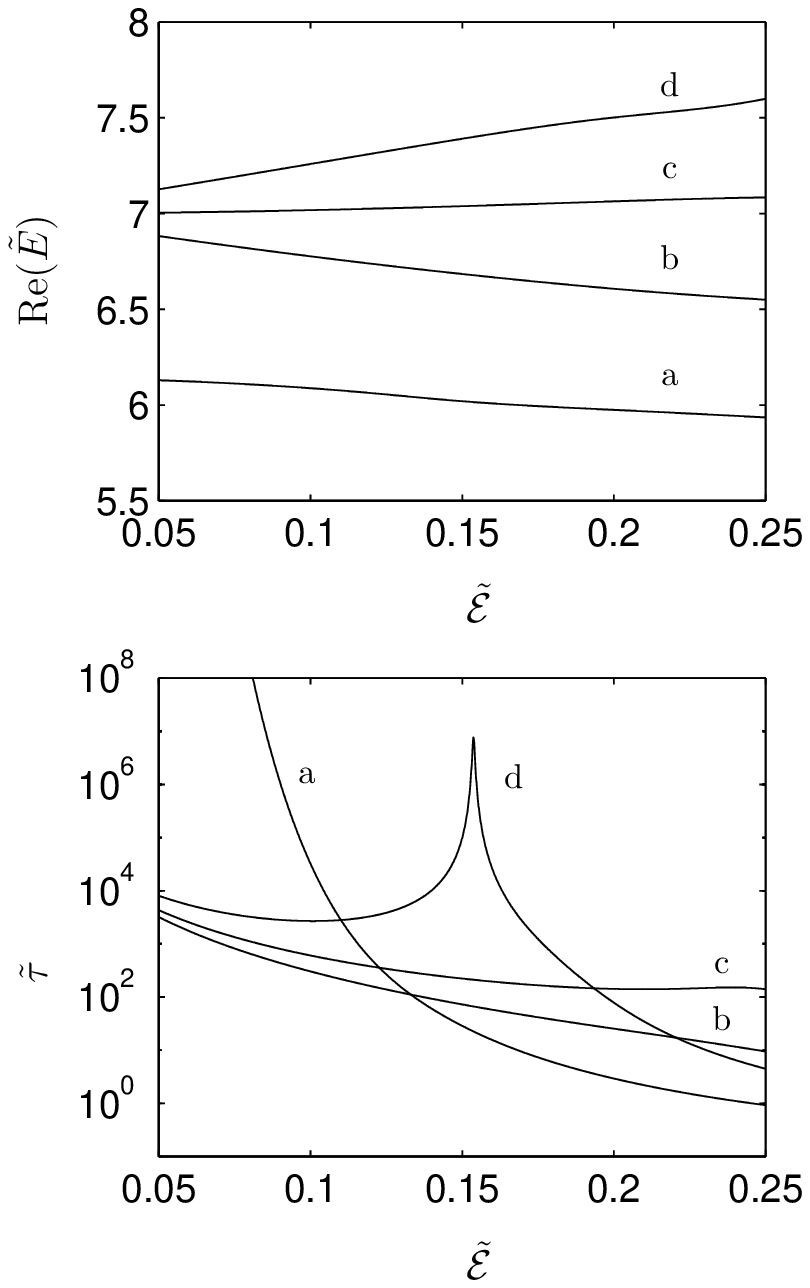}
\end{center}
\caption{Real parts and lifetimes of some resonance states as the
function of $\tilde{\cal E}$ for $\tilde{E}_B=-3$.}
\label{fig6}
\end{figure}

\section{Resonance states in external fields}
\label{sec:resonance}

In order to determine resonance energies for a non-vanishing
electric field we have to find complex zeros of $D(\tilde{E})$
defined by eq. (\ref{e3.4}) and lying below the real axis. This
means that we have to solve one complex equation,
$D(\tilde{E})=0$, with four unknown real parameters:
$\text{Re}(\tilde{E})$, $\text{Im}(\tilde{E})$, $\tilde{E}_B$
and the scaled electric field $\tilde{\cal{E}}$. To this end
we shall fix one of these parameters, treat another one as a free
parameter, and find the remaining two parameters by solving this
equation, i.e., by determining poles of the full Green's function.
At the beginning let us fix the imaginary part of the resonance
energy and find all possible zeros of $D(\tilde{E})$. The results
are shown in Figs. \ref{fig2} and \ref{fig3}. In Fig. \ref{fig2}
we present the real part, $\text{Re}(\tilde{E})$, of all
resonance states of a given $\text{Im}(\tilde{E})=10^{-4}$ for
the bound state energy $\tilde{E}_B$ between $-30$ and 0, as it
has been already presented in Fig. \ref{fig1} for the vanishing
electric field (in both these figures we show only those states
for which $\text{Re}(\tilde{E})>1$, i.e., lying above the first
Landau level). We observe that, apart from the already existing
impurity-induced bound states for $\tilde{\cal{E}}=0$, a
non-vanishing electric field generates new resonance states which
are located in the vicinity of Landau levels of energy
$\hbar\omega (n+\frac{1}{2})$ for $n$ greater than 0. Moreover, as
it follows from our numerical investigations, the number of these
new resonance states is equal to $n$. This means that for a
non-vanishing electric field and for a given quantum system,
specified by the binding energy $\tilde{E}_B$, the number of
impurity-induced resonances per Landau level is equal to $n+1$,
where the quantum number $n$ labels the Landau energy. It appears
that for the vanishing electric field real parts of energies of
these new resonances approach the Landau energy, as it is shown in
Fig. \ref{fig3}, in which we present the real part,
$\text{Re}(\tilde{E})$, of all resonance states of a given
$\text{Im}(\tilde{E})=10^{-4}$ as the function of
$\tilde{\cal{E}}$. We observe that for a given imaginary part
of resonances there is a maximum electric field above which no
such resonance states exist. This is not a surprising fact. What
is unexpected, however, is that the maximum electric field very
little depends on the imaginary part of resonance energies chosen,
as it is shown in Fig. \ref{fig4} for
$\text{Im}(\tilde{E})=10^{-6}$; although the lifetime of these
\textit{new states} has increased hundred times the maximum
electric field has decreased only by a small fraction. This
finding could suggest that even for nonperturbative electric
fields there could exist long-living resonance states, as it will
be discussed in details in \cite{KK2}.

The lines in Figs. \ref{fig3} and \ref{fig4} corresponding to the
electric-field-induced resonances are, in fact, a narrow loops. In
Fig. \ref{fig5} we present in an enlarged scale two such
resonances labelled by $\#1$ and $\#2$ in Fig. \ref{fig3}. It is
clearly seen that for these particular resonances paths of the
real part of resonance energies with a changing electric field
represent closed loops which start from a Landau level. For a
fixed electric field there are two states which correspond to two
different binding energies $\tilde{E}_B$ and, as could be
expected, the lower parts of these loops correspond to the
stronger attraction, i.e., to the larger values of
$|\tilde{E}_B|$.

\begin{table} 
\caption{Physical values of the
magnetic field $\cal{B}$, electric field $\cal{E}$ and
lifetime $\tau$ for some chosen values of the bound state energy
$E_B$ for a particular resonance.}
\label{tab1}
\begin{ruledtabular}
\begin{center}
\begin{tabular}{cccc}
 $E_B$ [meV] & $B$ [T] & $\cal{E}$ [kV/m] & $\tau$ [ns]\\ \hline
 1&0.506&6.423&7.52\\
 2&1.013&18.167&3.76\\
 4&2.025&51.383&1.88\\
 6&3.038&94.396&1.25 
\end{tabular}
\end{center}
\end{ruledtabular}
\end{table}

In order to estimate the lifetime of one of these new resonance
states let us take $\tilde{\cal{E}}=0.2647$ and
$\tilde{E}_B=-2.2860459726451$ for which there exists the
resonance of energy $\tilde{E}=3.0703456182811-10^{-4}\text{i}$.
This resonance corresponds in Fig. \ref{fig3} to the rightmost
point lying near the first excited Landau level. In Table
\ref{tab1} we show some numerical values of the magnetic field
$\cal{B}$ (in teslas), electric field $\cal{E}$ (in
kilovolts per meter) and lifetime $\tau$ (in nanoseconds) for this
resonance state and for some typical values of the binding energy
$E_B$ of an electron with effective mass $m^*=0.067$ (in the
electron's mass unit) trapped by an attractive impurity in the
GaAs quantum well \cite{PL90}. Our evaluations show that in spite
of applying strong electric fields, which, in general, are
expected to destabilize a system, one can observe generation of
long-living states with lifetimes comparable to or even larger
than typical timescales encountered in semiconductor
heterostructures.

Let us close this paper discussing the resonances extracted by an
attractive impurity and electric field from the third excited
Landau level of $n=3$. We fix now the binding energy by assuming
that $\tilde{E}_B=-3$ and draw the dependence of the complex
resonance energy $\tilde{E}$ on the electric field intensity
$\tilde{\cal{E}}$. As we have shown this above, the number of
such states is equal to $n+1$ and $n$ of them approach the $n$-th
Landau level for the vanishing electric field. The same conclusion
can be reached looking at the upper panel of Fig. \ref{fig6},
while in the lower panel we present lifetimes of these states. We
see that the state 'a', related to the old states discussed before
in \cite{P81,CC98}, is very stable for weak electric fields, but
with an increasing $\tilde{\cal{E}}$ its lifetime suddenly
decreases. On the contrary, the lifetimes of new
electric-field-induced resonances, discussed in this paper, are
shorter for weak fields, but with the increasing
$\tilde{\cal{E}}$ we observe a peculiar behavior for one of
them, i.e., for the resonance 'd'; instead of a continuous
decrease we observe that for some particular values of
$\tilde{\cal{E}}$ the lifetime increases, approaches its
maximum value, and then starts decreasing. We call this phenomenon
the electric-field-induced stabilization and we relate it to
quantum vortices, positions of which are controlled by an electric
field. A detailed analysis of this problem will be presented in
our subsequent paper \cite{KK2}.

\acknowledgments
This work has been supported in part by the Polish Committee
for Scientific Research (Grant No. KBN 2 P03B 039 19).


\begin{thebibliography}{99}
\bibitem{P81}
R. E. Prange, Phys. Rev. B 23 (1981) 4802.
\bibitem{PJ82}
R. E. Prange, R. Joynt, Phys. Rev. B 25 (1982) 2943.
\bibitem{JP84}
R. Joynt, R. E. Prange, Phys. Rev. B 29 (1984) 3303.
\bibitem{PC91}
F. Perez, F. A. B. Coutinho, Am. J. Phys. 59 (1991) 52.
\bibitem{CC98}
R. M. Cavalcanti, C. A. A. de Carvalho, J. Phys. A 31 (1998) 2391.
\bibitem{GM99}
S. Gyger, P. A. Martin, J. Math. Phys. 40 (1999) 3275.
\bibitem{HL99}
E. H. Hauge, J. M. J. van Leeuwen, Physica A 268 (1999) 525.
\bibitem{AGHH88}
S. Albeverio, F. Gesztesy, R. H\o egh-Krohn, H. Holden, Solvable
Models in Quantum Mechanics, Springer, Heidelberg, 1988.
\bibitem{B75}
I. J. Berson, J. Phys. B 8 (1975) 3078.
\bibitem{FFR90}
F. H. M. Faisal, P. Filipowicz, K. Rz\c a\.{z}ewski, Phys. Rev. A
(1990) 6176.
\bibitem{FFR91}
P. Filipowicz, F. H. M. Faisal, K. Rz\c a\.{z}ewski, Phys. Rev. A
(1991) 2210.
\bibitem{MFSF00}
N. L. Manakov, M. V. Frolov, A. F. Starace, I. I. Fabrikant, J.
Phys. B 33 (2000) R141.
\bibitem{W91}
K. W\'odkiewicz, Phys. Rev. A 43 (1991) 68.
\bibitem{IZ80}
C. Itzykson, J. B. Zuber, Quantum Field Theory, McGraw-Hill, New
York, 1980.
\bibitem{BBCK}
I. Bia\l ynicki-Birula, M. Cieplak, J. Kami\'nski, Theory of
Quanta, Oxford University Press, New York, 1992.
\bibitem{KK2}
K. Krajewska, J. Z. Kami\'nski, in preparation.
\bibitem{PL90}
T. Pang, S. G. Louie, Phys. Rev. Lett. 65 (1990) 1635.
\end{thebibliography}
\end{document}